\documentstyle[twoside,12pt]{article}
\normalsize
\newcommand{\bdi}{\begin{displaymath}}
\newcommand{\edi}{\end{displaymath}}
\newcommand{\bfi}{\begin{figure}}
\newcommand{\efi}{\end{figure}}

\newcommand{\beq}{\begin{equation}}
\newcommand{\eeq}{\end{equation}}

\newcommand{\gam}{\gamma_{\mu}}

\newcommand{\gaf}{\gamma_{5}}
\newcommand{\ra}{\rightarrow}
\newcommand{\dsla}{\partial\hspace{-6pt} /  }  
\newcommand{\Asla}{A\hspace{-6.5pt}  /  } 
 
\newcommand{\wt}{\widetilde}

\begin{document}
\begin{titlepage}
\begin{flushright}
\today
\end{flushright}

\vspace{1cm}
\begin{center}
{\Large \bf Theta vacuum in different gauges}\\[1cm]
C. Adam* \\
School of Mathematics, Trinity College, Dublin 2 \\

\vfill
{\bf Abstract} \\
\end{center}

In some recent papers it is claimed that the physical significance of
the vacuum angle theta for QCD-like theories depends on the chosen
gauge condition. We criticise the arguments that were given in support
of this claim, and show by explicit construction for the case of QED$_2$
that and why they fail, confirming thereby the commonly accepted point
of view.

\vfill

$^*)${\footnotesize  
email address: adam@maths.tcd.ie, adam@pap.univie.ac.at}
\end{titlepage}

\section{Introduction}

In gauge theories where large gauge transformations (i.e. gauge 
transformations with a nonzero, integer winding number) exist, a new
parameter, the vacuum angle $\theta$, enters the quantized theory, which
is absent classically. The commonly accepted point of view is that this
vacuum angle $\theta$ has physical consequences both for a pure
Yang-Mills theory and for a QCD-like theory provided all the quarks
are massive. 

However, in some recent papers \cite{Yaza1,Yaza2,Yama1} (see also 
\cite{Ki1}) this
point of view was criticised. There it was argued that the physical
relevance or irrelevance of $\theta$ depends on the chosen quantization
prescription. More precisely, it was claimed that whenever the Gauss
law defining the space of physical states,
\beq
G'[\lambda] |{\rm phys}\rangle =0 \quad {\rm or} \quad e^{iG[\lambda]}
|{\rm phys}\rangle \sim |{\rm phys}\rangle
\eeq
(here $\lambda$ is a gauge function, and $G$ and $G'$ are related), 
implements invariance w.r.t. small
gauge transformations only, physics does not depend on $\theta$. In the
opposite case, when gauge invariance w.r.t. both small and large gauge
transformations is implemented via (1), physics depends on $\theta$.

As a consequence, in all covariant quantization prescriptions (like e.g. BRST
quantization), where only ``small'' gauge invariance is implemented, 
physics should be independent of $\theta$. E.g., there would be no strong
CP violation for $\theta\ne 0$, and the topological susceptibility
would be zero.

In this paper first we briefly review and criticise the arguments that
were presented by the authors of \cite{Yaza1,Yaza2,Yama1} in favour of
their point of view. Both their arguments are quite general, so they 
should apply, e.g., to QED$_2$ as well as to QCD$_4$.

Therefore, we study the problem of the $\theta$ vacuum explicitly for 
pure two-dimensional electro-dynamics and for the massless and massive 
Schwinger models. We show that and why both arguments of
\cite{Yaza1,Yaza2,Yama1} fail and confirm the commonly accepted point
of view that physics depends on $\theta$ for pure electro-dynamics and for 
the massive Schwinger model, whereas it is independent of $\theta$ in 
the massless Schwinger model.

\section{Criticism of the arguments}

In \cite{Yaza1,Yaza2,Yama1} the following two arguments are given:
\begin{enumerate}

\item The topological charge density $\nu (x)$ is the divergence of a 
topological current $K_\mu$, $\nu =\partial_\mu K^\mu$. The topological 
charge  operator $Q_T =\int d^{d-1}x K_0$ of this current relates different
$\theta$ vacua,
\beq
|\theta\rangle =e^{i\theta Q_T}|0\rangle
\eeq
as well as different Hamiltonians $H_\theta$ (with corresponding 
Lagrangians $L_\theta$)
\beq
H_\theta =e^{i\theta Q_T}H_0 e^{-i\theta Q_T} \quad ,\quad
L_\theta =L_0 +\int d^{d-1}x \theta \nu .
\eeq
Therefore, the theory $H_\theta$, $|\theta\rangle$ is unitarily equivalent
to the theory $H_0$, $|0\rangle$.

This argument fails when ``large'' gauge invariance is implemented in (1),
because then the unitary operator $\exp (i\theta Q_T)$ does not commute 
with the Gauss operator $G[\lambda]$.

\item The VEV of the topological current is a constant in a covariant gauge,
\beq
\langle \theta |K_\mu (x)|\theta\rangle = \langle \theta |K_\mu (0)
|\theta\rangle ={\rm const.}
\eeq
(where translation invariance of $|\theta\rangle$ is used) and, therefore,
the topological density $\nu$ has zero VEV, $\partial_\mu \langle\theta
|K^\mu (x)|\theta\rangle = \langle\theta |\nu (x)|\theta\rangle =0$.
Further, the vacuum energy density $\epsilon$ is independent of $\theta$
because of $\langle\theta |\nu (x)|\theta\rangle \sim \frac{\partial 
\epsilon}{\partial \theta}$.

The current $K_\mu$ is gauge variant, but as the vacuum state $|\theta\rangle$
is a proper (normalized) state of the original, ``large'' Hilbert space
(the one without a physical state condition that contains pure gauge
excitations as well), the VEV $\langle\theta |K_\mu |\theta\rangle$ 
nevertheless exists.
\end{enumerate}
Our criticism of these two arguments is as follows.
\begin{enumerate}
\item[ad 1)]
It is, of course, true that the two theories $H_0$, $|0\rangle$ and
$H_\theta$, $|\theta\rangle$ are equivalent, but this is not the way the 
$\theta$ parameter enters the theory. Instead, it relies on the fact that, 
for a given theory $H_0$, the Gauss law (1) 
allows for a one-parameter family of states that may be parametrized by
$\theta$. Therefore, the $\theta$-dependent theory is $H_0$, $|\theta\rangle$, 
which may be written equivalently as $H_\theta$, $|0\rangle$ due to (2), (3).
\item[ad 2)]
The assumption on the existence of $\langle\theta |K_\mu |\theta\rangle$ is 
incorrect. A $|\theta\rangle$ state can only be constructed after the Gauss 
law (1) is implemented. This may be done
e.g. by solving the Gauss law (see Section 3) or by explicitly constructing
$|\theta\rangle$ states out of the gauge invariant (i.e. Gauss-law obeying)
subspace of the conventional, perturbative Fock space (see Section 4). In
both cases the Gauss law has to be implemented {\em before} a $|\theta\rangle$ 
state can be constructed. Therefore, the VEV $\langle\theta |K_\mu 
|\theta\rangle$ does not exist.
\end{enumerate}
In the following two sections we shall demonstrate our arguments by
explicit construction, first in pure two-dimensional electro-dynamics, 
then in the Schwinger model.

\section{Pure electro-dynamics}

We use the conventions $\eta_{00}=-\eta_{11}=1$, $\epsilon_{01}=1$.
The Lagrangian density for two-dimensional electro-dynamics with a 
$\theta$ term reads ($F_{01}\equiv F$)
\beq
{\cal L} =-\frac{1}{4} F_{\mu\nu}F^{\mu\nu} -\frac{e\theta}{4\pi}
\epsilon_{\mu\nu}F^{\mu\nu}
=\frac{1}{2}F^2 +\frac{e\theta}{2\pi}F
\eeq
\beq
F_{\mu\nu}=\partial_\mu A_\nu -\partial_\nu A_\mu .
\eeq
First of all, as ${\cal L}$ only depends on $F$, we may, in principle, treat
$F$ as the canonical field. Then we find the equation of motion
(for the ``Heisenberg field'' $F$)
\beq
F=-\frac{e\theta}{2\pi}
\eeq
and the Hamiltonian density
\beq
{\cal H}=-{\cal L}.
\eeq
As the field $F$ is equal to a (c-number) constant, the theory is very
trivial. There is only one physical state, the vacuum $|0\rangle$
(more precisely, $F$ does not generate new states out of the vacuum) and
$F$ and ${\cal H}$ act on $|0\rangle$ like
\beq
F|0\rangle =-\frac{e\theta}{2\pi}|0\rangle \quad ,\quad
{\cal H}|0\rangle =\frac{1}{2}(\frac{e\theta}{2\pi})^2 |0\rangle .
\eeq
Next, let us reanalyse the theory (5) when treating $A_\mu$ as the 
canonical field. In this case the equation of motion reads
\beq
\partial_\mu F=0,
\eeq
i.e., again, $F$ is an (unspecified) constant.

Further, we want to define the topological density $\nu$, current $K_\mu$
and charge $Q_T$,
\beq
\nu =\frac{e}{4\pi}\epsilon_{\mu\nu}F^{\mu\nu}=-\frac{e}{2\pi}F_{01}
\eeq
\beq
K_\mu =\frac{e}{2\pi}\epsilon_{\mu\nu}A^\nu
\eeq
\beq
Q_T  =-\frac{e}{2\pi}\int dx A_1 (x)
\eeq
and the unitary operator
\beq
U(\alpha) =e^{i\alpha Q_T}
\eeq
($x\equiv x^1$).
We shall discuss the theory within canonical quantization and use the gauge
condition $A_0 =0$. Further, we rename the $\theta$ parameter in ${\cal
L}$ into $\theta'$. Then the Lagrangian density, canonical momentum $\Pi$
of the field $A_1$ and Hamiltonian density read
\beq
{\cal L}_{\theta'} =\frac{1}{2}(\partial_0 A_1)^2 +\frac{e\theta'}{2\pi}
\partial_0 A_1 
\eeq
\beq
\Pi =-\partial_0 A_1 -\frac{e\theta'}{2\pi}
\eeq
\beq
{\cal H}_{\theta'} =\frac{1}{2}(\Pi +\frac{e\theta'}{2\pi})^2 .
\eeq
In $A_0 =0$ gauge there remains the residual gauge freedom of time-independent
gauge transformations $\lambda (x)$,
$A_1 (x)\rightarrow A_1 (x) +\frac{1}{e}\lambda' (x)$, which is implemented 
by the Gauss operator $G[\lambda]$ ($\lambda' \equiv \partial_1 \lambda$),
\beq
G[\lambda] =\frac{1}{e}\int dx\lambda' (x)\Pi (x)
\eeq
\beq
e^{iG[\lambda]}A_1 (x)e^{-iG[\lambda]}=A_1 (x) +\lambda' (x).
\eeq
Physical states should respect this symmetry, i.e., they should be invariant
under $\exp (iG[\lambda])$, up to a c-number phase \cite{JR1,CDG},
\beq
e^{iG[\lambda]}|{\rm phys}\rangle =e^{i\rho [\lambda]}|{\rm phys}\rangle ,
\eeq
where $\rho$ is a real number that may depend on $\lambda$. In fact, any
state on which the equation of motion (10) holds, obeys 
this Gauss law constraint.
Let us define
\beq
\Pi |\theta\rangle =-\frac{e\theta}{2\pi}|\theta\rangle
\eeq
then we get
\beq
e^{iG[\lambda]}|\theta\rangle =e^{-i\frac{\theta}{2\pi}w[\lambda]}
|\theta\rangle
\eeq
where
\beq
w[\lambda]=\lambda (\infty) - \lambda (-\infty)
\eeq
and each $|\theta\rangle$ is a physical state. Obviously, the Gauss law
(20) implies the more common local version of the Gauss law
\beq
\partial_1 \Pi |{\rm phys}\rangle =0,
\eeq
and both versions are equivalent on eigenstates of $\Pi$, (21).
The eigenvalue of the Hamiltonian density (17) is
\beq
{\cal H}_{\theta'}|\theta\rangle =\frac{1}{2}(\frac{e(\theta' 
-\theta)}{2\pi})^2 |\theta\rangle ,
\eeq
therefore, here the combination $(\theta' -\theta)$ is equivalent to
$\theta$ in (9) (we could have chosen $\theta' =0$ to make the equivalence
even more obvious). Clearly, different $|\theta\rangle$ states are orthogonal.
As $\exp (iG[\lambda])$ commutes with all the gauge invariant observables
($\Pi$, ${\cal H}_{\theta'}$), we may say that $\exp (iG[\lambda])$
defines a superselection rule, where different $|\theta\rangle$ states
correspond to different sectors of the theory. Each sector contains
precisely one state $|\theta\rangle$, the vacuum state in the given
$\theta$ sector.

Observe that nothing so far depended on the boundary conditions of the
gauge transformations $\lambda$, i.e., whether $w[\lambda]$ in (23) is
arbitrary, $w[\lambda] =2\pi m,\, m\in {\cal Z}$ (large gauge
transformations), or $w[\lambda]=0$ (small gauge transformations). These
restrictions just affect the value of the phase in (22), i.e.
$\exp (iG[\lambda])$ has the same eigenvalue 1 for all $\theta$ if
$w[\lambda] =0$, and $\exp (iG[\lambda]) $ has the same eigenvalue on
$|\theta\rangle$ and on $|\theta +2\pi\rangle$ if $w[\lambda]=2\pi$. 
Nevertheless, $\Pi$ and ${\cal H}_{\theta'}$ distinguish different
$|\theta\rangle$ states, independently of these boundary conditions. 
In addition, $\Pi$ and ${\cal H}_{\theta'}$ are invariant under gauge
transformations $\lambda$ with arbitrary $w[\lambda]$, therefore there
is no good reason for a restriction on $\lambda$.

The operator $U$, (14), maps different $|\theta\rangle$ states into each
other, 
\beq
|\theta +\theta' \rangle =U(\theta' )|\theta\rangle ,
\eeq
therefore we may use the operator $U$ to shift the $\theta$ dependence 
entirely to the Hamiltonian density via
\beq
{\cal H}_{\theta' -\theta}=U^\dagger (\theta){\cal H}_{\theta'} U(\theta)
=\frac{1}{2}(\Pi +\frac{e(\theta' -\theta)}{2\pi})^2
\eeq
\beq
{\cal H}_{\theta' -\theta}|0\rangle =U^\dagger (\theta){\cal H}_{\theta'} 
|\theta\rangle
=\frac{1}{2}(\frac{e(\theta' -\theta)}{2\pi})^2 |0\rangle .
\eeq
Of course, it holds that
\beq
{\cal H}_\theta |\theta\rangle =U (\theta) {\cal H}_0 U^\dagger (\theta)
U (\theta)|0\rangle =0,
\eeq
but this does not imply the physical equivalence of different 
$|\theta\rangle$ states.

Next, let us comment on  argument 2) concerning the VEV of the
topological current $K_\mu$. In our case $K_\mu \sim \epsilon_{\mu\nu}A^\nu$, 
i.e. after
gauge fixing $A_0 = 0$ we should study the VEV of $A_1$. For the 
introduction of the $|\theta\rangle$ states gauge invariance 
was essential, i.e. all $|\theta\rangle$
states obey the physical state condition (20).
However, this physical state condition is incompatible with the existence
of the VEV $\langle \theta |A_1 |\theta\rangle$, and, therefore,
such a VEV does not exist.

So we have found for our theory that ``physics'' depends on $\theta$ and
that different values of $\theta$ lead to different vacuum energy
densities for all $\theta \in {\cal R}$. This conforms with the physical
interpretation of $\theta$ as it was discussed in \cite{Co1}, where
$\theta$ is interpreted as a background static electrical field $E=
\frac{e\theta}{2\pi}$. As nothing can screen this field (there is no matter),
it may acquire arbitrary values.

Before we end this section, we want to show how a $2\pi$ periodicity in
$\theta$ may enter the theory, getting thereby also a clearer
understanding of the role of the unitary operator U, (14).
Usually, the $2\pi$ periodicity of $\theta$ is a consequence of the fact
that only gauge fields with integer instanton numbers contribute to VEVs
of physical observables. In our case obviously nothing restricts 
$F=\partial_0 A_1$ (we choose $A_0 =0$ gauge again) to integer instanton 
numbers (in fact, in the Euclidean version of the theory $F$ may be
interpreted as a magnetic field perpendicular to the Euclidean plane, and
this magnetic field may have arbitrary magnetic flux, which is proportional
to the ``instanton number'' of $F$, see \cite{Fry}). However, we may
introduce this restriction as a constraint.  
This is done most conveniently
in the path integral formalism, 
\begin{displaymath}
Z=\sum_{n=-\infty}^\infty \int DF\delta (\int d^2 x\frac{F}{2\pi}-en)
e^{i\int d^2 x(\frac{1}{2}F^2 +\frac{e\theta}{2\pi}F)}
\end{displaymath}
\beq
=\sum_{n=-\infty}^\infty \int DF e^{i\int d^2 x(\frac{1}{2}F^2 +
(\frac{\theta}{2\pi}+n)eF)},
\eeq
and in the second line the $2\pi$ periodicity of $\theta$ is obvious.
So we have a whole ensemble of systems, numerated by $n$, and the Lagrangian 
${\cal L}_n$ of system $n$ has an effective $\theta$ parameter 
$\theta' =\theta +2\pi n$. The corresponding Hamiltonian density is
${\cal H}_n =\frac{1}{2}(\Pi +\frac{e\theta}{2\pi}+en)^2$, see (17),
provided that all the $\theta$ dependence is shifted to ${\cal H}_n$,
i.e., the vacuum is $|0\rangle$.

The vacuum energy density of the whole ensemble of ${\cal H}_n$ is 
(remember that $\Pi |0\rangle =0$; $V$ $\ldots$ space-time volume)
\beq
\epsilon =-\frac{1}{V}\ln \Bigl( \sum_n e^{-\frac{e^2 V}{2}(\frac{\theta}{2\pi}
+n)^2}\Bigr) .
\eeq
In the thermodynamic limit $V\to \infty$ in the sum within the logarithm
only one term contributes, namely the one with $|\frac{\theta}{2\pi} +n|
={\rm min}$. As $\theta$ varies, the only contributing ${\cal H}_n$ 
changes whenever a point $|\theta | =\pi +2\pi m,\, m\in {\cal Z}$, is
passed. This makes the vacuum energy density $2\pi$ periodic in $\theta$
and equal to
\beq
\epsilon =\frac{1}{2}(\frac{e(\theta +2\pi n)}{2\pi})^2 \quad {\rm for}
\quad -\pi -2\pi n <\theta <\pi -2\pi n ,
\eeq
and the contribution solely stems from ${\cal H}_n$ (similar arguments
can be found in \cite{Sm1}, \cite{HaZh}).

Here an important point is that on the whole ensemble $\sum_n \exp (-\int
d x {\cal H}_n)$, the operator $U(2\pi)$ acts as a symmetry, although an
individual ${\cal H}_n$ is not symmetric under the action of $U(2\pi)$.
As $U(2\pi)$ maps $U(2\pi)|\theta\rangle =|\theta +2\pi \rangle$, this is
another way of stating the $2\pi$ periodicity in $\theta$.

Observe that the whole discussion had nothing to do with the boundary
conditions of gauge transformations. Still, each ${\cal H}_n$ in the
ensemble is invariant under gauge transformations with arbitrary
``winding number'' $w[\lambda]/2\pi$, and there is no need to restrict, 
e.g., to integer winding number.

\section{Massless and massive QED$_2$}

In the following we choose Lorentz gauge $\partial_\mu A^\mu =0$
(which is covariant), because
there the wellknown operator solution \cite{LS1,AAR,Belv} is at hand.
In fact, our presentation will closely follow Chapter 10 of
\cite{AAR}.  
We use the $\gamma$-matrix conventions
\beq
\gamma^{0} =\left( \begin{array}{cc}0 & 1 \\ 1 & 0 \end{array} \right)
\; ,\qquad
\gamma^{1} =\left( \begin{array}{cc}0 & -1 \\ 1 & 0 \end{array} \right)
\; ,\qquad
\gaf =\left( \begin{array}{cc}1 & 0 \\ 0 & -1 \end{array} \right) ,
\eeq
and by $c$ we always denote an unspecified constant whose precise value
we do not need.
The Lagrangian density of one-flavour QED$_2$ is
\beq
{\cal L}= \bar\Psi (i\dsla +e\Asla )\Psi -\frac{1}{4}F_{\mu\nu}F^{\mu\nu}
-m\bar\Psi \Psi ,
\eeq
but we want to consider the massless case $m=0$ first. The quantum equations 
of motion are (in the following we bosonize the theory, and normal ordering 
w.r.t. the bosonic fields is always understood)
\beq
i\dsla \Psi +e\Asla \Psi =0
\eeq
\beq
\partial_\mu F^{\mu\nu} +eJ^\nu =0
\eeq
\beq
\partial_\mu J_5^\mu =\frac{e}{2\pi}\epsilon_{\mu\nu}F^{\mu\nu},
\eeq
where (37) is the anomaly equation ($J_5^\mu =\epsilon^{\mu\nu}J_\nu$). In
Lorentz gauge $A_\mu$ may be written as
\beq
A_\mu =-\frac{\sqrt{\pi}}{e}(\epsilon_{\mu\nu}\partial^\nu \Sigma
+\partial_\mu \wt \eta )
\eeq
\beq
F_{\mu\nu} =\frac{\sqrt{\pi}}{e}\epsilon_{\mu\nu}\Box \Sigma ,
\eeq
where $\wt \eta$ is a massless, free scalar field ($\Box \wt\eta =0$) that
takes into account the residual gauge freedom (remember that each free,
massless scalar field $\eta$ has a dual $\wt\eta$ via $\partial_\mu \eta =
\epsilon_{\mu\nu}\partial^\nu \wt\eta$). 

The Dirac equation is solved by
\beq
\Psi =ce^{i\sqrt{\pi}\gaf (\Sigma +\eta)}\psi^{\rm f}
\eeq
where $\psi^{\rm f}$ solves the free Dirac equation and reads
\beq
\psi^{\rm f} =ce^{i\sqrt{\pi}(\wt\phi +\gaf \phi)}\psi_0 .
\eeq
Here $\psi_0$ is the constant unit spinor, $\psi_0^{\rm t} =(1,1)$,  
$\phi$ is another free, massless field and $\wt\phi$ is its dual.

From the Maxwell and anomaly equations we find $\Box (\Box +e^2 /\pi)
\Sigma =0$, i.e., $\Sigma$ is the sum of a free massless and a free massive
(with Schwinger mass $e/\sqrt{\pi}$) field. Choosing $\Sigma$ to be 
purely a massive field, $(\Box +e^2 /\pi)\Sigma =0$ (which is
possible), gives the following results: the vector current is  
\beq
J_\mu =-\frac{1}{\sqrt{\pi}}\epsilon_{\mu\nu}\partial^\nu \Sigma + L_\mu
\quad ,\quad L_\mu := -\frac{1}{\sqrt{\pi}}\epsilon_{\mu\nu}
\partial^\nu (\phi +\eta ),
\eeq
where the free, massless field $\phi$ (41) is the prepotential of the free 
fermion 
current, $j^{\rm f}_\mu =\bar \psi^{\rm f}\gam \psi^{\rm f}=-(1/\sqrt{\pi})
\epsilon_{\mu\nu}\partial^\nu \phi$. The Maxwell equation reads
\beq
\epsilon_{\mu\nu}\partial^\nu (\Box +\frac{e^2}{\pi})\Sigma -\frac{e^2}{
\sqrt{\pi}}L_\mu
\eeq
and cannot be fulfilled on the operator level. Instead, the space of physical 
states has to be defined via
\beq
\langle {\rm phys}|L_\mu |{\rm phys'}\rangle =0,
\eeq
and all operators $O$ leaving the space of physical states invariant,
i.e. $\langle {\rm phys}|L_\mu O|{\rm phys'}\rangle =0$, may be chosen
as observables.
In addition, the theory may be equivalently represented by the bosonic
Hamiltonian density
\beq
{\cal H}_{\rm bos}=\frac{1}{2}\Bigl( (\partial_0 \Sigma)^2 +(\partial_1
\Sigma)^2 +\frac{e^2}{\pi}\Sigma^2 +(\partial_0 \phi)^2 +(\partial_1
\phi)^2 -(\partial_0 \eta)^2 -(\partial_1 \eta)^2 \Bigr) ,
\eeq
where $\eta$ is quantized with opposite sign, $[\eta (t,x),\partial_0
\eta (t,y)]=-i\delta (x-y)$, in order to maintain $\langle 0|[L_\mu (x),
L_\nu (y)]|0\rangle =0$, which follows from the physical state condition.
Here $|0\rangle$ is the Fock vacuum of the fields $\Sigma$, $\eta$ and $\phi$
(which certainly is a physical state w.r.t. condition (44); 
for details see \cite{AAR}).

{\em Remark}: shifting $\Sigma$ by a constant, $\Sigma \ra \Sigma +c$, 
obviously is a symmetry of the underlying theory, see (38), (40), (41)
(in fact, this shift just generates global chiral rotations
$\Psi \ra \exp (i\sqrt{\pi}\gaf c)\Psi$). However, this manifest symmetry is 
lost in (45) due to the choice of $\Sigma$ as a purely massive field. 
Therefore, whenever we create a constant shift of $\Sigma$ by some
symmetry transformation, we should compensate it by a redefinition of
the $\Sigma$ field, $\Sigma' =\Sigma +c$.

Now the construction of the $\theta$ vacuum proceeds as follows. The
operator
\beq
\chi =e^{2\sqrt{\pi}i (\phi +\eta)}
\eeq
is an observable and commutes with the Hamiltonian density 
${\cal H}_{\rm bos}$. Therefore, it may be used to generate an infinite 
set of ground states out of the Fock vacuum $|0\rangle$,
\beq
|n\rangle =\chi^n |0\rangle .
\eeq
On physical states $\chi$ acts like a unitary implementer of large gauge
transformations (with winding number one, see \cite{AAR} and below), 
and we allow
only for integer powers of $\chi$ in (47) (see below for a more detailed
explanation of this point). The $|\theta\rangle$ vacuum
is just the coherent superposition of these $n$-vacua, and $\chi$ is diagonal 
w.r.t. $|\theta\rangle$,
\beq
|\theta\rangle =\sum_n e^{-in\theta}|n\rangle \quad ,\quad 
\chi |\theta\rangle =e^{i\theta}|\theta\rangle .
\eeq
Now we should study the operators 
\beq
U(\alpha)= e^{i\alpha Q_T}=e^{-i\frac{\alpha}{2\sqrt{\pi}}\int dx
(\partial_0 \Sigma +\partial_0 \eta)}
\eeq
\beq
V(\alpha)= e^{-i\alpha Q_5}=e^{i\frac{\alpha}{2\sqrt{\pi}}
\int dx\partial_0 \phi},
\eeq
where $Q_5$ is the chiral charge, $Q_5 =1/2 \int dx j^{\rm f}_{5,0}$
(we choose the factor 1/2 in order to have one unit of chiral charge for
the chiral density $\bar \Psi (1/2) ({\bf 1}+\gaf )\Psi$). Both operators 
associate the same charge $n$ to the $n$-vacuum $|n\rangle$,
\beq
(\chi^\dagger)^n U(\alpha)\chi^n =e^{in\alpha}U(\alpha)
\eeq
\beq
(\chi^\dagger)^n V(\alpha)\chi^n =e^{in\alpha}V(\alpha) ,
\eeq
which demonstrates the fact that ``instanton number equals chirality''.

Both operators transform the $|\theta\rangle$ vacuum,
\beq
U(\theta)|\theta'\rangle =V(\theta)|\theta'\rangle =|\theta +\theta'
\rangle .
\eeq
In addition, both operators leave the Hamiltonian ${\cal H}_{\rm bos}$
invariant. For $V(\alpha)$ this is obvious, $V(\alpha){\cal H}_{\rm bos}
V^\dagger (\alpha)={\cal H}_{\rm bos}$. $U(\alpha)$ shifts the field $\Sigma$ 
in ${\cal H}_{\rm bos}$ to $\Sigma -\alpha$. However, as we remarked,
this is due to a convention that hides the symmetry of the original theory
under such a shift and should be compensated by a redefinition
$\Sigma \ra \Sigma -\alpha$.

Therefore, we conclude that the theory $H_\theta$, $|0\rangle$ is
equivalent to the theory $H_0$, $|0\rangle$ in this case and, consequently,
physics is independent of $\theta$ in the massless Schwinger model
(here $|0\rangle$ denotes the $\theta$ vacuum for $\theta =0$, not the
Fock vacuum).

Now, let us briefly discuss the massive Schwinger model. The vacuum
structure remains the same, but the bosonized Hamiltonian density (45)
changes to
\beq
{\cal H}_m ={\cal H}_{\rm bos}+cm\cos\Bigl( 2\sqrt{\pi} (\Sigma +\phi
+\eta )\Bigr) .
\eeq
Therefore, ${\cal H}_m$ is no longer invariant under the action of
$V(\theta)$ or $U(\theta)$, but instead
\beq
V(\theta){\cal H}_m V^\dagger (\theta)=U(\theta){\cal H}_m U^\dagger (\theta)
={\cal H}_{\rm bos} +cm\cos \Bigl( 2\sqrt{\pi}(\Sigma +\phi +\eta )
+\theta \Bigr)
\eeq
(where again a shift $\Sigma \ra \Sigma -\theta$ has to be performed when
the action of $U(\theta)$ is computed).

As a consequence, $H_\theta$, $|0\rangle$ and $H_0$, $|0\rangle$ are no
longer equivalent and the $\theta$ dependence enters the massive
Schwinger model in the wellknown fashion \cite{CJS,Co1,AAR,MSMPT,MSSM}.

\smallskip


Concerning the existence of a VEV $\langle\theta |K_\mu |\theta\rangle$, 
we should observe that the vacuum raising operator $\chi$, (46), acts as
a constant operator only on gauge invariant operators. Therefore, a VEV
of a gauge variant quantity like $K_\mu$ w.r.t. a $|n\rangle$ state is
arbitrary and not translation invariant. Consequently, $K_\mu$ does not
have a VEV w.r.t. the $|\theta\rangle$ vacuum.

\smallskip

We have presented our main arguments, but we want to add some comments before
closing this section. In (44) the longitudinal vector $L_\mu$ vanishes
only weakly, which has the advantage that the physical states just form 
a subset of the entire Fock space (of free fields for the massless Schwinger
model; however, the combination $\eta +\phi$ remains a free, massless field
even for nonzero fermion mass). As can be shown easily, $L_\mu$ generates zero
norm states when applied to the Fock vacuum (or any other physical state).
These zero norm states are redundant on gauge invariant observables, so
one may define a new space of physical states by just dividing out the
zero norm states. On this new space the physical state condition 
corresponding to (44) is
\beq
L_\mu |{\rm phys}\rangle =0,
\eeq 
which is similar to the local version (24) of the Gauss law in Section 3.
At this point the question arises whether an integrated version of the
Gauss law, analogous to (20), can be found, which could again give some insight
into the role of winding gauge transformations. Actually, this may be done 
without problem. The Gauss operator $G[\lambda]$ that implements the residual
gauge transformations on $\Psi$ and $A_\mu$,
\beq
e^{iG[\lambda]}\Psi (x)e^{-iG[\lambda]}=e^{i\lambda (x)}\Psi (x)
\eeq
\beq
e^{iG[\lambda]}A_\mu (x)e^{-iG[\lambda]}=A_\mu (x)+\frac{1}{e}\partial_\mu
\lambda (x) ,
\eeq
is given by
\beq
G[\lambda]=\frac{1}{\sqrt{\pi}}\int dx^1 \Bigl( (\eta (x) +\phi (x))
\partial_1 \lambda (x) -(\wt\eta (x) +\wt\phi (x))\partial_0 \lambda (x)\Bigr)
\eeq
as may be checked easily (see \cite{AAR}). Remember that, because of the
Lorentz gauge $\partial_\mu A^\mu =0$, the residual gauge transformations
$\lambda$ have to obey $\Box \lambda =0$. Therefore, $\lambda$ may be
decomposed like
\beq
\lambda (x)= \lambda_+ (x^1 +x^0) +\lambda_- (x^1 -x^0) ,
\eeq
and the Gauss operator may be rewritten, by using this decomposition and by 
some partial integrations, like
\bdi
G[\lambda]= \int dx^1 \Bigl( -(\lambda_+ +\lambda_- )L_0 -(\lambda_+ -
\lambda_- )L_1 \Bigr) +
\edi
\beq
\frac{1}{\sqrt{\pi}}\Bigl( (\eta +\phi)(w[\lambda_+] +w[\lambda_-])
-(\wt\eta +\wt\phi)(w[\lambda_+] -w[\lambda_-])\Bigr)
\eeq
where
\beq
w[\lambda_\pm]=\lambda_\pm (\infty) -\lambda_\pm (-\infty) ,
\eeq
and we used the fact that $(\eta +\phi)$ (and $(\wt\eta +\wt\phi)$) acts as a
constant operator on all gauge invariant operators, and, consequently,
on physical states. $L_0$ and $L_1$ are just the components of the 
longitudinal current $L_\mu$, see (42), (56). 
At this point we want to impose the restriction that
the total bare electrical charge is zero,
\beq
\int dx^1 j^{\rm f}_0 (x)=\int dx^1 \partial_1 \phi (x)=0.
\eeq
This assumption is very reasonable because of the confining properties of
the Schwinger model, and we already implicitly obeyed it by considering only
uncharged operators as observables. (However, this assumption is by no
means necessary; it is perfectly possible to include charged sectors and
observables into the theory, which just somewhat complicates the vacuum
structure, see \cite{LS1,AAR}.) For the zero charge condition (63) to be
respected by a gauge transformation (61), we have to assume that
\beq
w[\lambda_+]-w[\lambda_-] =0 \quad \Rightarrow \quad 
w[\lambda_+] =w[\lambda_-] =\frac{1}{2}w[\lambda] .
\eeq
In this case the Gauss operator simplifies to   
\beq
G[\lambda]=-\int dx^1 \lambda^\mu L_\mu +\frac{1}{\sqrt{\pi}}
w[\lambda](\eta +\phi) ,
\eeq
where we introduced the short-hand notation $\lambda^0 :=\lambda_+ +\lambda_-$,
$\lambda^1 :=\lambda_+ -\lambda_-$. 

Now it is very easy to write down the integrated version of the Gauss law
analogous to (20),
\beq
e^{iG[\lambda]}|{\rm phys}\rangle =e^{i\rho[\lambda]}|{\rm phys}\rangle .
\eeq
Choosing
\beq
e^{2\sqrt{\pi}i(\eta +\phi)}|\theta\rangle \equiv \chi |\theta\rangle
=e^{i\theta}|\theta\rangle
\eeq
as in (48), we find
\beq
e^{iG[\lambda]}|\theta\rangle =e^{i\frac{\theta}{2\pi}w[\lambda]}|\theta\rangle
\eeq
(see (22)), and, again, the integrated version, (66), of the Gauss law implies 
the local one, (56).

From the preceding discussion it is very clear where the $\theta$ dependence
and $2\pi$ periodicity come from: if we restrict the set of 
observables to operators
with zero chirality (i.e. operators that do not contain $\chi$) we find no
$\theta$ dependence (like for the Hamiltonian of the massless Schwinger model);
if we restrict to operators with integer chiral charge (in our units (50)),
i.e., operators containing integer powers of $\chi$ (like $\bar\Psi 
(1/2)({\bf 1}+\gaf)\Psi =c\chi\exp(2\sqrt{\pi}i\Sigma)$), we find $2\pi$
periodicity in $\theta$. Of course, this latter restriction is very
reasonable from a physical point of view. The important point is that all
these results have nothing to do with a possible restriction of the 
residual gauge transformations $\lambda$
(e.g. to integer or zero winding number).

What is affected by restrictions on $\lambda$ is the number of physical states
that obey the integrated Gauss law (66). For unrestricted $\lambda$ the
$|\theta\rangle$ states (67) are the only allowed states (of course, 
toghether with their excitations by the physical, massive $\Sigma$ field),
and the Gauss law (66) defines a superselection rule. 
For integer winding number
$w[\lambda]=2\pi m$ linear combinations $\sum_n a_n |\theta +2\pi n\rangle$
also obey the Gauss law (66), and for zero winding number $w[\lambda]=0$
arbitrary linear superpositions of $|\theta\rangle$ states obey the Gauss law
(66), and the integrated and local versions of the Gauss law become
equivalent. Such restrictions on $\lambda$, of course, cannot change the
physical results we described so far. E.g., even if we allow for the
Fock vacuum to be a physical state, it cannot be the vacuum state of the
massive Schwinger model, because it is not even an eigenstate of the
Hamiltonian ${\cal H}_m$. On the other hand, the $|\theta\rangle$ vacuum
always belongs to the set of physical states, and VEVs of observables
w.r.t. $|\theta\rangle$ do not depend on what restrictions are imposed
on the residual gauge transformations $\lambda$. In addition, as already
stated, there is no good reason to restrict $\lambda$, because all
gauge invariant observables are invariant under gauge transformations
$\exp (iG[\lambda])$ for arbitrary $\lambda$.

\section{Summary}

We have criticised the arguments of \cite{Yaza1,Yaza2,Yama1} by explicit
construction of the simplest models where they should apply, and
demonstrated their failure, which was the purpose of this article.
We showed that the implementation of invariance for arbitrary, large or
small gauge transformations via the Gauss law (1) just affects the
number of states that fulfill the Gauss law. The $|\theta\rangle$ vacuum
always obeys the Gauss law, and VEVs of observables w.r.t. $|\theta\rangle$
do, of course, not depend on the number of further states that are
permitted by the Gauss law. We confirmed that the vacuum energy densities
of pure two-dimensional electro-dynamics and of the massive Schwinger
model depend on $\theta$, whereas the vacuum energy density of the massless
Schwinger model is independent of $\theta$, as expected.

\section*{Acknowledgement}

The author thanks the members of the Department of Mathematics at Trinity
College, where this work was performed, for their hospitality. Further
thanks are due to R. Jackiw for helpful comments.


\begin{thebibliography}{999999}
\bibitem{Yaza1}
H. Yamagishi, I. Zahed, hep-th 9709125
\bibitem{Yaza2}
H. Yamagishi, I. Zahed, hep-ph 9507296
\bibitem{Yama1}
H. Yamagishi, Prog. Theor. Phys. 87 (1992) 785
\bibitem{Ki1}
H. Kikuchi, Int. J. Mod. Phys. A9 (1994) 2741 
\bibitem{JR1}
R. Jackiw, C. Rebbi, Phys. Rev. Lett. 37 (1976) 172
\bibitem{CDG}
C. Callan, R. Dashen, D. Gross, Phys. Lett. B63 (1976) 334
\bibitem{Co1}
S. Coleman,  Ann. Phys.  101 (1976) 239
\bibitem{Fry}
M. P. Fry, Phys. Rev. D45 (1992) 682, D47 (1993) 2629 
\bibitem{Sm1}
A. V. Smilga, Phys. Rev. D54 (1996) 7757
\bibitem{HaZh}
I. Halperin and A. Zhitnitsky, hep-ph 9711398
\bibitem{MSMPT}
C. Adam, Ann. Phys. 259 (1997) 1
\bibitem{LS1}
J. Lowenstein, J. Swieca,  Ann. Phys.  68 (1971) 172
\bibitem{AAR}
E. Abdalla, M. Abdalla, K. D. Rothe, "2 dimensional Quantum Field Theory",
World Scientific, Singapore 1991
\bibitem{Belv}
L. V. Belvedere, K. D. Rothe, B. Schroer, J. A. Swieca, Nucl. Phys.
153 (1979) 112
\bibitem{CJS}
S. Coleman, R. Jackiw, L. Susskind,  Ann. Phys.  93 (1975) 267
\bibitem{MSSM}
C. Adam,  Phys. Lett.  B 363 (1995) 79
\end{thebibliography}
\end{document}